\documentclass[preprint,a4paper,superscriptaddress,floatfix]{revtex4-2}

\usepackage[left=13.5mm,right=13.5mm,top=17mm,bottom=34mm]{geometry}

\usepackage{graphicx}
\usepackage{xcolor}
\usepackage{amsmath}
\usepackage{amsfonts}
\usepackage{amssymb}
\usepackage{bm}
\usepackage{upgreek}

\usepackage{listings}
\usepackage{soul}

\usepackage{array}
\newcommand{\PreserveBackslash}[1]{\let\temp=\\#1\let\\=\temp}
\newcolumntype{C}[1]{>{\PreserveBackslash\centering}p{#1}}
\newcolumntype{R}[1]{>{\PreserveBackslash\raggedleft}p{#1}}
\newcolumntype{L}[1]{>{\PreserveBackslash\raggedright}p{#1}}

\usepackage[
unicode=true,
bookmarks=true,
bookmarksnumbered=true,
hyperfootnotes=false,
colorlinks=true,
linkcolor=blue,
citecolor=blue,
urlcolor=blue
]{hyperref}

\newenvironment{mat}{\left[\begin{array}{ccccccccccccccc}}{\end{array}\right]}
\newcommand\bcm{\begin{mat}}
\newcommand\ecm{\end{mat}}

\newenvironment{cmat}{\left(\begin{array}{ccccccccccccccc}}{\end{array}\right)}
\newcommand\bcrm{\begin{cmat}}
\newcommand\ecrm{\end{cmat}}

\newenvironment{rmat}{\left[\begin{array}{rrrrrrrrrrrrr}}{\end{array}\right]}
\newcommand\brm{\begin{rmat}} 
\newcommand\erm{\end{rmat}}

\begin{document}

%TC:ignore

\title{\texorpdfstring{
\begin{center}
{Dynamic bifurcation of Resch origami}\\
{for self-adaptive impact mitigation}
\end{center}
}{Dynamic bifurcation of Resch-patterned origami for self-adaptive impact mitigation}}

\author{Yasuhiro Miyazawa}
\affiliation{Department of Mechanical Engineering, Seoul National University, 1 Gwanak-ro, Gwanak-gu, Seoul 08826, South Korea}
\affiliation{Department of Aeronautics and Astronautics, University of Washington, Seattle, Washington  98195-2400, USA}

\author{Dahun Lee}
\affiliation{Department of Mechanical Engineering, Seoul National University, 1 Gwanak-ro, Gwanak-gu, Seoul 08826, South Korea}

\author{Seonghyun Kim}
\affiliation{Department of Mechanical Engineering, Seoul National University, 1 Gwanak-ro, Gwanak-gu, Seoul 08826, South Korea}

\author{Chia-Yung Chang}
\affiliation{Department of Aeronautics and Astronautics, University of Washington, Seattle, Washington  98195-2400, USA}

\author{Qixun Li}
\affiliation{Department of Aeronautics and Astronautics, University of Washington, Seattle, Washington  98195-2400, USA}

\author{Ryan Tenu Ahn}
\affiliation{Department of Chemistry, University of Washington, Seattle, Washington  98195-2400, USA}

\author{Minho Cha}
\affiliation{Department of Mechanical Engineering, Seoul National University, 1 Gwanak-ro, Gwanak-gu, Seoul 08826, South Korea}

\author{Koshiro Yamaguchi}
\affiliation{Department of Mechanical Engineering, Seoul National University, 1 Gwanak-ro, Gwanak-gu, Seoul 08826, South Korea}
\affiliation{Department of Aeronautics and Astronautics, University of Washington, Seattle, Washington  98195-2400, USA}

\author{Yuyang Song}
\affiliation{Toyota Research Institute North America, 1555 Woodridge Ave, Ann Arbor, Michigan 48105, USA}

\author{Shinnosuke Shimokawa}
\affiliation{Toyota Research Institute North America, 1555 Woodridge Ave, Ann Arbor, Michigan 48105, USA}

\author{Umesh Gandhi}
\affiliation{Toyota Research Institute North America, 1555 Woodridge Ave, Ann Arbor, Michigan 48105, USA}

\author{Jinkyu Yang}
\email[]{jkyang11@snu.ac.kr}
\affiliation{Department of Mechanical Engineering, Seoul National University, 1 Gwanak-ro, Gwanak-gu, Seoul 08826, South Korea}

\date{\today}

\maketitle
%TC:endignore

\section*{Abstract}\label{abstract}
A long-standing challenge in impact mitigation is the development of versatile and omnifarious protective structures capable of encompassing a wide spectrum of scenarios, for example, ranging from low-speed pedestrian impacts to high-speed vehicle collisions.
However, most existing impact mitigation strategies rely on fixed geometries or pre-tuned material properties targeting specific impact speeds, lacking the ability to adapt in real time~\cite{Davoodi2012, Zheng2014, Meza2014, Shan2015, Yuan2019, Wang2019, Bastek2023, Fancher2023, Smith2024, Chao2025}.
Here, we introduce a Resch-patterned origami structure~\cite{Resch1970, Tachi2013, Yang2022, Yu2023} that hosts two distinct deformation modes: a monostable folding mode and a bistable unfolding mode featuring snap-through.
{Impact experiments reveal a dynamic bifurcation, wherein the incoming impact velocity determines which of the two deformation pathways is activated.
Unlike kinematic or static origami bifurcations governed by geometric compatibility or imposed displacement~\mbox{\cite{Ario2010, Silverberg2015Origami, Waitukaitis2015, Yasuda2017, Gillman2018, Sadeghi2020, Ma2021, Tomita2023MatDes, Li2024}}, this dynamic bifurcation enables the structure to engage distinct energy-dissipation mechanisms under various impact conditions.}
We further demonstrate the scalability and practical relevance of this mechanism by fabricating tessellations in a bumper-like configuration and evaluating their performance using a pendulum-based mannequin impact test. 
These results establish dynamic bifurcation in origami-based structures as a strategy for scalable and programmable impact-mitigation systems that autonomously select deformation modes in real time, with potential applications in adaptive robotics, protective armor, and aerospace structures.

\clearpage

Origami principles have been proven effective in designing deployable and shape-morphing structures, leading to the exploration of origami-based mechanical metamaterials~\cite{Lang2016, Kim2019, Lee2021, Meng2022}.
These systems exhibit properties such as load-bearing capability~\cite{Schenk2013, Saito2014, Yasuda2015, Yang2018, Melancon2021, Jamalimehr2022, Zhang2023, Zhu2024}, variable stiffness~\cite{Mukhopadhyay2020, Miyazawa2021}, auxeticity~\cite{Yasuda2015, Zheng2022}, and multi-stability~\cite{Silverberg2014, Yang2018, Novelino2020, Wang2024}, which are highly tunable through geometric modifications.
{Extending this programmability to target dynamic impact is particularly attractive for impact-mitigation applications, where structures must accommodate a broad range of loading conditions.}
{Dynamic responses of origami-based structures have been investigated in one-dimensional nonlinear lattices~\mbox{\cite{Yasuda2019}}, tubular absorbers~\mbox{\cite{Qiang2024}}, stacked or volumetric cellular architectures~\mbox{\cite{Kshad2019, Harris2021, Tomita2023, Yang2023}}, and sheet-like tessellations~\mbox{\cite{Dong2021, Xia2022}}.
Yet, realizing a spatially extended origami surface that can efficiently dissipate impact energy remains an important challenge.}

\begin{figure}[tb]
    \centering
    \includegraphics[width=\linewidth]{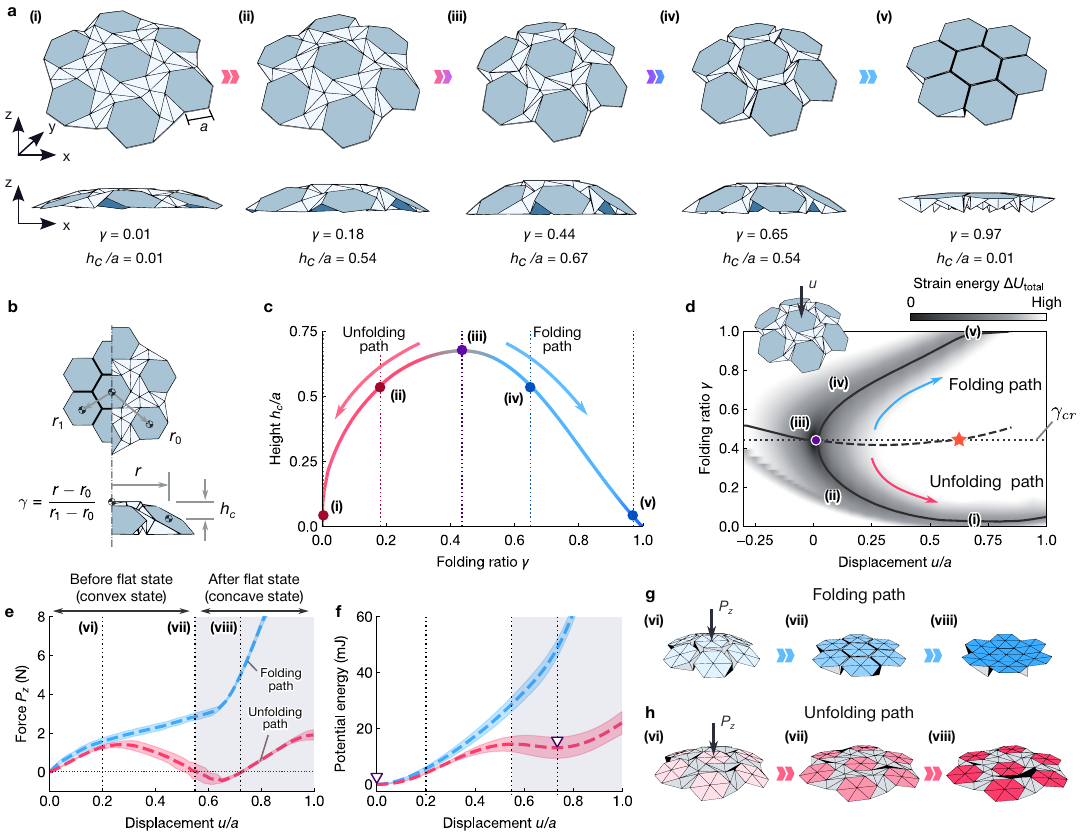}
    \caption{\textbf{Hexagonal Resch origami and its folding motion.}
    (\textbf{a}) Top and side views of the Resch origami folding.
    The dimension of the hexagonal Resch pattern is determined solely by the side length of the hexagon $a$.
    (\textbf{b}) Definition of height $h_{\rm c}$ and folding ratio $\gamma$.
    (\textbf{c}) Variation of normalized height $h_{\rm c}/a$ as a function of folding ratio $\gamma$.
    (\textbf{d}) {Strain} energy landscape and bifurcation diagram under out-of-plane {compression $u$}.
    {The grayscale color intensity represents the strain energy stored in the bar-and-hinge model, comprising the torsional-spring energy associated with crease folding and the axial-bar energy associated with panel deformation (Supplementary Note~4).}
    {The Roman numerals in (\textbf{c}, \textbf{d}) correspond to the configurations in (\textbf{a}).
    In (\textbf{d}), the purple circle and orange star denote the bifurcation point and the mode-selection threshold, respectively.}
    (\textbf{e}) Experimentally measured compressive force \(P_z\) as a function of displacement for the folding (blue) and unfolding (red) configurations of the PET specimens.
    (\textbf{f}) Corresponding potential-energy profiles obtained by integrating the measured force with respect to displacement. {Triangles indicate the local minima (stable states) along the unfolding path.}
        Postures during deformation along
    (\textbf{g}) folding and
    (\textbf{h}) unfolding path.
    }
    \label{fig:main_schematic}
\end{figure}

{Here, we use Resch-patterned origami to show that an extended, surface-forming structure can autonomously select its deformation pathway according to the impact speed.
Originally developed for versatile architectural surfaces~\mbox{\cite{Resch1970}}, Resch patterns transition between flat and convex configurations during folding, enabling free-form curvature programming~\mbox{\cite{Tachi2013, Yang2022, Yu2023}}.}
The Resch origami considered in this study consists of hexagonal panels connected by triangular folding units termed startucks~\cite{Tachi2013}.
In particular, we focus on a circular tiling pattern, in which the hexagonal facets are tessellated radially while preserving rotational symmetry~(Fig.~\ref{fig:main_schematic}a).
During folding, the structure undergoes radial contraction accompanied by rotation of each hexagonal facet about its centroid (Supplementary Video~1).
It evolves from an initially flat state through a convex configuration and ultimately reaches a fully folded flat state (Fig.~\ref{fig:main_schematic}a, side view)~\cite{Resch1970, Tachi2013}.

We characterize the folding sequence using two morphometric quantities, surface height $h_{\rm c}$ and folding ratio $\gamma$~(Fig.~\ref{fig:main_schematic}b).
As $\gamma$ increases from $0$ to $1$, $h_{\rm c}$ initially rises monotonically, reaching a maximum~($h_{\rm c, max}\approx0.67a$) at the critical folding ratio $\gamma_{\rm cr}\approx0.44$~(Fig.~\ref{fig:main_schematic}c).
Beyond $\gamma_{\rm cr}$, the height decreases, returning to zero when the Resch pattern is fully folded.
This characterization of the folding motion reveals two distinctive deformation paths from the most convex configuration ({purple dot labeled (iii)}): a folding path (blue arrow) and an unfolding path (red), naturally exhibiting intrinsic bifurcation behavior.

The strain energy landscape~(Fig.~\ref{fig:main_schematic}d) indeed identifies a supercritical pitchfork bifurcation, where a single stable path (solid line) splits into two stable paths under compression ($u/a>0$), separated by an unstable equilibrium branch (dashed line).
The stable paths at $u/a>0$ correspond to the height-folding ratio relationship in Fig.~\ref{fig:main_schematic}c, confirming that the dual-mode kinematic behavior arises under compression. To characterize these responses experimentally, we perform quasi-static compression tests on polyethylene terephthalate (PET) Resch specimens preconfigured at 80\% of the maximum height~($h_{\mathrm{c},0}=0.8h_{\mathrm{c,max}}$; Fig.~1a,c, configurations~(ii) and~(iv)), corresponding to the folding and unfolding configurations, respectively.

Along the folding path, the force increases monotonically with weak strain-softening (Fig.~\ref{fig:main_schematic}e).
{Facet contact begins at $u/a\approx0.54$, after which the facets initially accommodate further compression through bending.
The force therefore rises sharply only at larger displacement, when densification becomes dominant~(Fig.~{\ref{fig:main_schematic}}g, (viii)).}
This results in a single stable state at $u=0$ in the potential energy profile~(Fig.~\ref{fig:main_schematic}f, triangle).
Contrarily, the unfolding path shows both positive and negative force regimes, giving rise to bistability as confirmed by the potential energy profile~(Fig.~\ref{fig:main_schematic}f), where two minima are observed.
Examining  the deformed postures reveals that, before reaching the second stable state, the Resch origami attains the fully deployed flat configuration~(Fig.~\ref{fig:main_schematic}h, (vii)).
Consequently, the second stable state extends beyond the flat state, undergoing further deformation with snap-through and eventually adopting a stable concave shape~(Fig.~\ref{fig:main_schematic}h, (viii)).
 See Methods, Supplementary Notes~1,~2, and Supplementary Videos~2,~3 for more details.

\section*{Dynamic bifurcation under variable impact conditions}

\begin{figure*}[htbp]
    \centering
    \includegraphics[width=\linewidth]{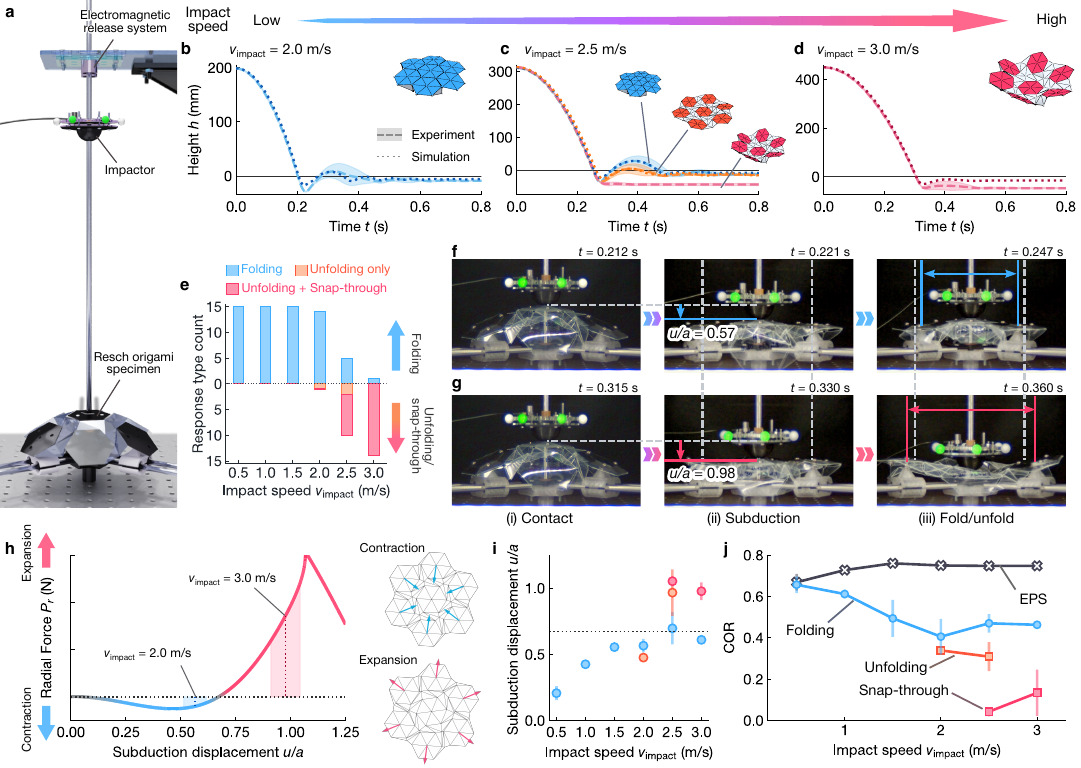}
    \caption{\textbf{Dynamic bifurcation of Resch origami.}
    (\textbf{a}) Drop tower configuration.
    {The drop tests use a hemispherical impactor with a diameter of $D_\mathrm{impactor}=32$~mm.}
    The height of the impactor at
    (\textbf{b}) $v_{\rm impact}=2.0$ m/s,
    (\textbf{c}) $2.5$ m/s, and
    (\textbf{d}) $3.0$ m/s.
    {Dashed lines, experimental mean; shaded regions, standard deviation over five trials; dotted lines, simulation.}
    The inset figures of each panel schematically display the folding mode that Resch origami follows.
    (\textbf{e}) Histogram of folding, unfolding, and snap-through as a function of impact speed.
    Deformation sequence of the Resch origami in the experiment undergoing (i) contact, (ii) subduction, and (iii) fold/unfold phases when
    (\textbf{f}) $v_\mathrm{impact}=2.0$ m/s and 
    (\textbf{g}) $3.0$ m/s.
    (\textbf{h}) Numerically estimated radial force exerted on the outer hexagons during the subduction phase.
    Sub-panels correspond to the subduction displacement in
    (top) $v_{\rm impact}=2.0$ and
    (bottom) $3.0$ m/s cases.
    (\textbf{i}) Subduction displacement as a function of impact speed.
    (\textbf{j}) Coefficient of restitution as a function of impact speed.
    Blue {circle} symbol, folding mode;
    orange square symbol, unfolding mode;
    red square symbol, snap-through mode;
    black cross symbol, EPS.
    }
    \label{fig:main_dynamic_bifurcation}
\end{figure*}

Building on the static bifurcation behavior, we conduct impact tests on the Resch origami pre-configured at the bifurcation configuration (Fig.~\ref{fig:main_schematic}c, (iii)), preconfigured at $\gamma\approx0.44$ and $h_{\rm c}=h_{\rm c, max}$.
The details of a drop test (Fig.~\ref{fig:main_dynamic_bifurcation}a) are provided in Methods, Supplementary Note~3 and Supplementary Video~4.
At a lower impact speed ($v_\mathrm{impact}=2.0$ m/s), the impactor rebounds back to $h>0$ after collision, producing an oscillatory height profile (Fig~\ref{fig:main_dynamic_bifurcation}b) and
indicating that potential energy stored in the folding motion is released after the impact.
At $v_{\rm impact}=2.5$ m/s, three distinct trajectories appear: (i) folding with an oscillatory rebound~(Fig.~\ref{fig:main_dynamic_bifurcation}c, blue), (ii) unfolding without reaching the second stable state~(orange), and (iii) snap-through, where the impactor {is abruptly brought to rest} as the structure settles into the second stable state~(red).
Further increasing the impact speed to $3.0$ m/s results exclusively in the snap-through response~(Fig.~\ref{fig:main_dynamic_bifurcation}d).

Figure~\ref{fig:main_dynamic_bifurcation}e quantifies this transition by classifying each trial of the drop test according to folding, unfolding, or snap-through and presenting the results as a histogram.
Folding dominates below $2.0$ m/s, whereas at $2.5$ m/s, five out of fifteen trials follow the folding path and ten follow the unfolding path, eight of which undergo snap-through.
At $3.0$ m/s, snap-through occurs in $93$\% of trials.
These results confirm that the Resch origami exhibits a speed-dependent transition between deformation modes, {which is distinct from previously reported kinematic or quasi-static origami bifurcations governed by geometric compatibility or imposed displacement~\mbox{\cite{Ario2010, Silverberg2015Origami, Waitukaitis2015, Yasuda2017, Gillman2018, Sadeghi2020, Ma2021, Tomita2023MatDes, Li2024}}.}

The deformation sequences at $v_\mathrm{impact}=2.0$ and $3.0$ m/s reveal the mechanism of real-time pathway selection~(Fig.~\ref{fig:main_dynamic_bifurcation}f,g for the folding and unfolding trials, respectively).
The impact initiates with a contact phase, followed by a subduction phase where the impactor continues to descend and drives the downward translation of the center hexagon (see horizontal dashed lines in Fig.~\ref{fig:main_dynamic_bifurcation}f,g).
During this subduction phase, the centroids of the outer hexagons coincide between the folding and unfolding trials (vertical dashed lines in the center panels), indicating negligible radial motion.
Only after substantial subduction does an activation of the radial contraction or expansion of the outer hexagons occur (vertical dashed lines, rightmost panels).
This delayed response of the outer hexagons originates from the highly compliant startuck connections linking the center and outer hexagons, which slow the transmission of the impact and postpone radial activation until significant central deformation occurs.

To explain the selective contraction-expansion mechanism, Fig.~\ref{fig:main_dynamic_bifurcation}h plots the radial force acting on outer hexagons against the normalized subduction displacement $u/a$ ({Extended Data Fig.~{\ref{fig:ex_radial_force_multiorbit}}}, Supplementary Note~4.9).
The direction of the radial force depends sensitively on the subduction displacement: for $u/a<0.67$, the force is inward (negative), favoring folding; for $u/a>0.67$, it becomes outward (positive), promoting unfolding.
Consequently, the Resch origami follows distinct deformation pathways depending on the attained subduction displacement, which is governed by the impact speed.
This threshold subduction displacement coincides with the maximum height of the Resch origami, indicating that the switching occurs near its flat configuration.
Moreover, this threshold approximately matches the point where the unstable branch crosses $\gamma_\mathrm{cr}$ (Fig.~\ref{fig:main_schematic}d, star), marking the switching between the folding- and unfolding-favored energy wells.
The unstable path initially leans toward the unfolding path from $\gamma_{\rm cr}$, and subsequently toward the folding path at larger subduction displacements.
Thus, during the subduction phase, the Resch origami preferentially falls into the folding (unfolding) energy well for small (large) subduction displacements.

Empirical data corroborate this trend.
At $v_{\rm impact}=2.0$ m/s, the average subduction is $u/a=0.57\pm0.051$, inducing an inward radial force and resulting in a folding mode.
Conversely, at $3.0$ m/s, a large subduction of $u/a=0.98\pm0.066$ induces an outward radial force, leading to snap-through.
Figure~\ref{fig:main_dynamic_bifurcation}i summarizes this dependence; the subduction displacement increases with impact speed, reaching the threshold value $u/a \approx 0.67$ at $v_\mathrm{impact} \approx 2.5$ m/s.
{These results show that increasing impact speed drives the central region beyond the radial force reversal point, switching the dynamic response between the folding and unfolding pathways identified in the quasi-static strain-energy landscape in Fig.~\mbox{\ref{fig:main_schematic}}.}

The dynamic bifurcation endows the Resch origami with enhanced impact-absorbing behavior compared to the conventional material (Fig.~\ref{fig:main_dynamic_bifurcation}j).
We quantify the performance using the coefficient of restitution (COR): 
$\left|\frac{v_{\rm r}}{v_{\rm impact}}\right|$, where $v_{\rm r}$ is the rebound velocity of the impactor.
In the folding mode (blue), the COR decreases with increasing impact speed, indicating enhanced energy absorption at higher velocities.
The transition from folding to unfolding~(blue to orange) leads to a further reduction in COR.
When snap-through occurs (red), the COR drops sharply, approaching near-zero restitution.
By contrast, an extruded polystyrene foam (EPS) specimen of comparable mass and projected area exhibits an approximately constant and relatively high COR across the same range of impact speed (black; Methods and Supplementary Note~1; {further comparison with conventional energy-absorbing materials provided in Supplementary Note~6.}).
These results demonstrate the adaptive impact response of the multi-modal Resch architecture over the tested speed range (Supplementary Notes~5).
{The stainless-steel specimen exhibits the corresponding speed-dependent transition at a higher energy scale (Extended Data Fig.~{\ref{fig:ex_dynamic_bifurcation_stainless}}), reflecting its greater stiffness and strain-energy capacity (Supplementary Notes~4.8 and 10.2).}

\section*{Scalability and feasibility tests in large-scale tessellations}

\begin{figure*}[h]
    \centering
    \includegraphics[width=\linewidth]{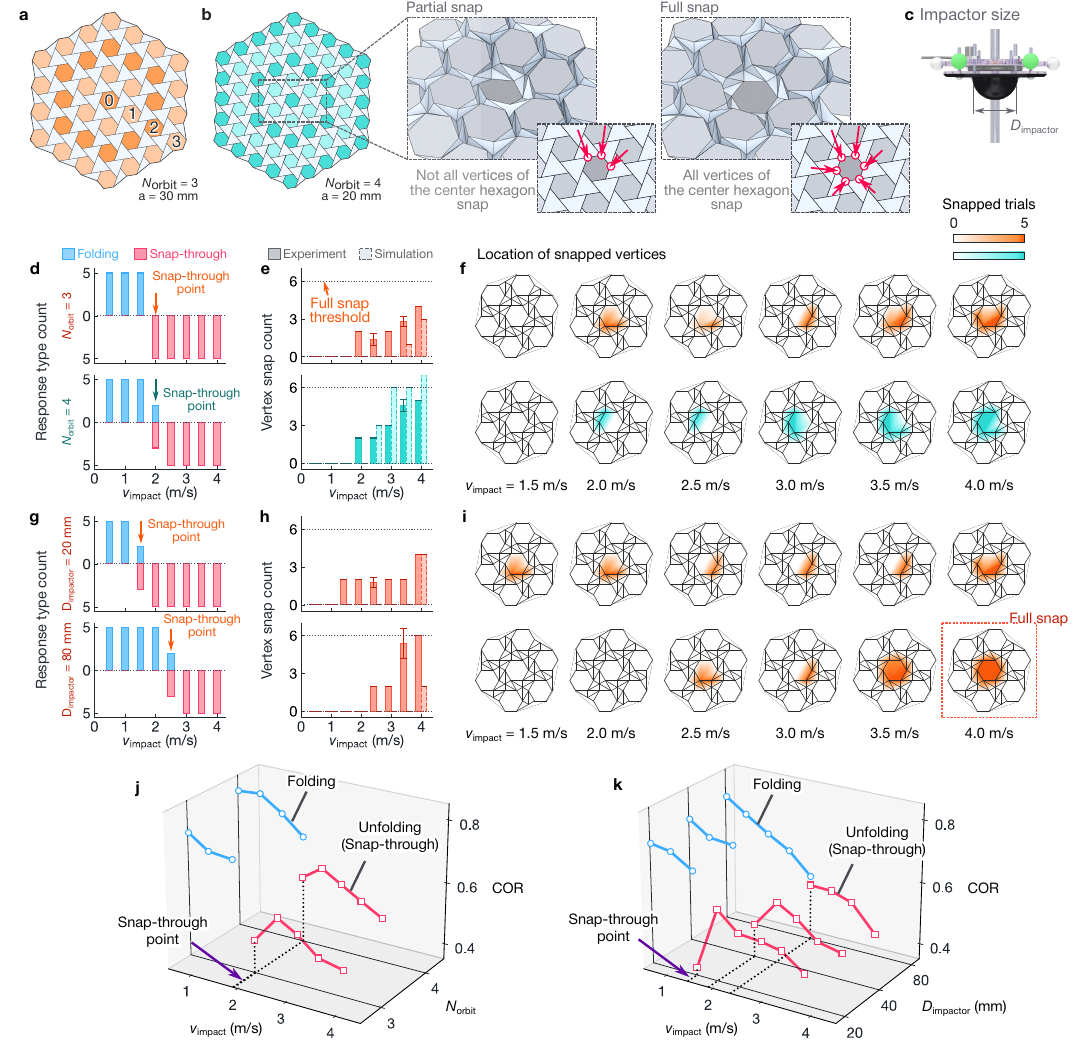}
    \caption{\textbf{Dynamic bifurcation of multi-orbit tessellation.}
    Crease patterns of
    (\textbf{a}) 3-orbit and
    (\textbf{b}) 4-orbit tessellations with the definition of partial and full snap of vertices.
    {The labels in panel (\textbf{a}) define the orbit-numbering convention:
    the central hexagon is denoted as orbit 0, and $N_\mathrm{orbit}$ counts the successive concentric orbits surrounding it.}
    (\textbf{c}) Impactor with a hemispherical head.
    Histograms of 
    (\textbf{d}) response type and  
    (\textbf{e}) number of snapped vertices, and  
    (\textbf{f}) {empirical} map of snapped vertex locations for the (top) 3- and (bottom) 4-orbit tessellations ($D_\mathrm{impactor}=40$ mm).
    (\textbf{g}-\textbf{i}) The same for the impactor size of (top) $D_\mathrm{impactor}=20$ mm and (bottom) $D_\mathrm{impactor}=80$ mm (3-orbit tessellation).
    {In panels (\textbf{f}) and (\textbf{i}), the shaded regions indicate snapped vertices; the color intensity denotes the number of trials (out of five) in which each vertex snapped, as indicated by the colorbars.
    Orange and cyan distinguish the 3- and 4-orbit tessellations, respectively.}
    (\textbf{j}) Effect of $N_\mathrm{orbit}$ with $D_\mathrm{impactor}=40$ mm.
    (\textbf{k}) Effect of $D_\mathrm{impactor}$ with $N_\mathrm{orbit}=3$.
    }
    \label{fig:main_dynamic_bifurcation_multiorbit}
\end{figure*}

We examine the generality and scalability of the dynamic bifurcation by performing drop tests on 3- and 4-orbit specimens (Fig.~{\ref{fig:main_dynamic_bifurcation_multiorbit}}a,b; Supplementary Note 7) and using impactors of different sizes $D_\mathrm{impactor}$ (Fig.~{\ref{fig:main_dynamic_bifurcation_multiorbit}}c).
To quantify the transition from folding to unfolding, we count the number of vertices undergoing snap-through~(side insets in Fig.~{\ref{fig:main_dynamic_bifurcation_multiorbit}}b).
{Here, the vertex snap count denotes the number of vertices on the hexagonal panels that undergo local snap-through during impact and serves as a measure of the effective capacity of the distributed energy reservoir activated by this local deformation; its relationship with energy absorption is further examined in Extended Data Fig.~{\ref{fig:ex_parametric_vertex_snap_count_cor}} and Supplementary Note~12.3.}
For $N_\mathrm{orbit} = 3$ and $D_\mathrm{impactor} = 40$ mm, snap-through first appears at $2.0$ m/s (top Fig.~{\ref{fig:main_dynamic_bifurcation_multiorbit}}d), and the number of snap-through vertices generally increases with impact speed (Fig.~{\ref{fig:main_dynamic_bifurcation_multiorbit}}e,f, top).
{The larger $N_\mathrm{orbit} = 4$ tessellation follows a similar trend, indicating that the localized deformation mechanism underlying the folding-to-unfolding transition persists as the tessellation is extended (Fig.~{\ref{fig:main_dynamic_bifurcation_multiorbit}}d--f, bottom).}

Reducing the impactor size to $D_\mathrm{impactor}=20$ mm further enhances this localization.
Snap-through occurs at a lower speed of $1.5$ m/s, triggering the earlier transition from folding to unfolding (Fig.~{\ref{fig:main_dynamic_bifurcation_multiorbit}}g--i, top).
In contrast, a larger impactor ($D_\mathrm{impactor}=80$ mm) delays the onset of snap-through due to a more spatially distributed interaction (Fig.~{\ref{fig:main_dynamic_bifurcation_multiorbit}}g--i, bottom).
Notably, although the snap-through is delayed with the larger impactor, once initiated, it progresses more rapidly to full snap of six vertices than with the smaller impactor case (Supplementary Note~7).

Figure~{\ref{fig:main_dynamic_bifurcation_multiorbit}}j--k summarizes the sensitivity of the COR to $N_\mathrm{orbit}$ and $D_\mathrm{impactor}$.
{The corresponding trends in both the COR and the snap-through point remain qualitatively similar across the experimentally tested orbit numbers, supporting the localized nature of the underlying mode-selection mechanism.}
By contrast, the snap-through point shifts with $D_\mathrm{impactor}$, identifying $D_\mathrm{impactor}$ as a practical control parameter for tuning the transition point.
{The numerical sweeps extend these experimental observations by mapping how the impactor diameter, mass, torsional stiffness, and initial folding ratio govern the critical speed and the number of snap-through vertices~(Extended Data Fig.~{\ref{fig:ex_parametric_sweep}}, Supplementary Note~13). 
This analysis identifies the parameter space governing both the onset and spatial extent of dynamic bifurcation, providing a framework for tuning the Resch structure to different impact conditions.}

\begin{figure*}[htbp]
    \centering
    \includegraphics[width=\linewidth]{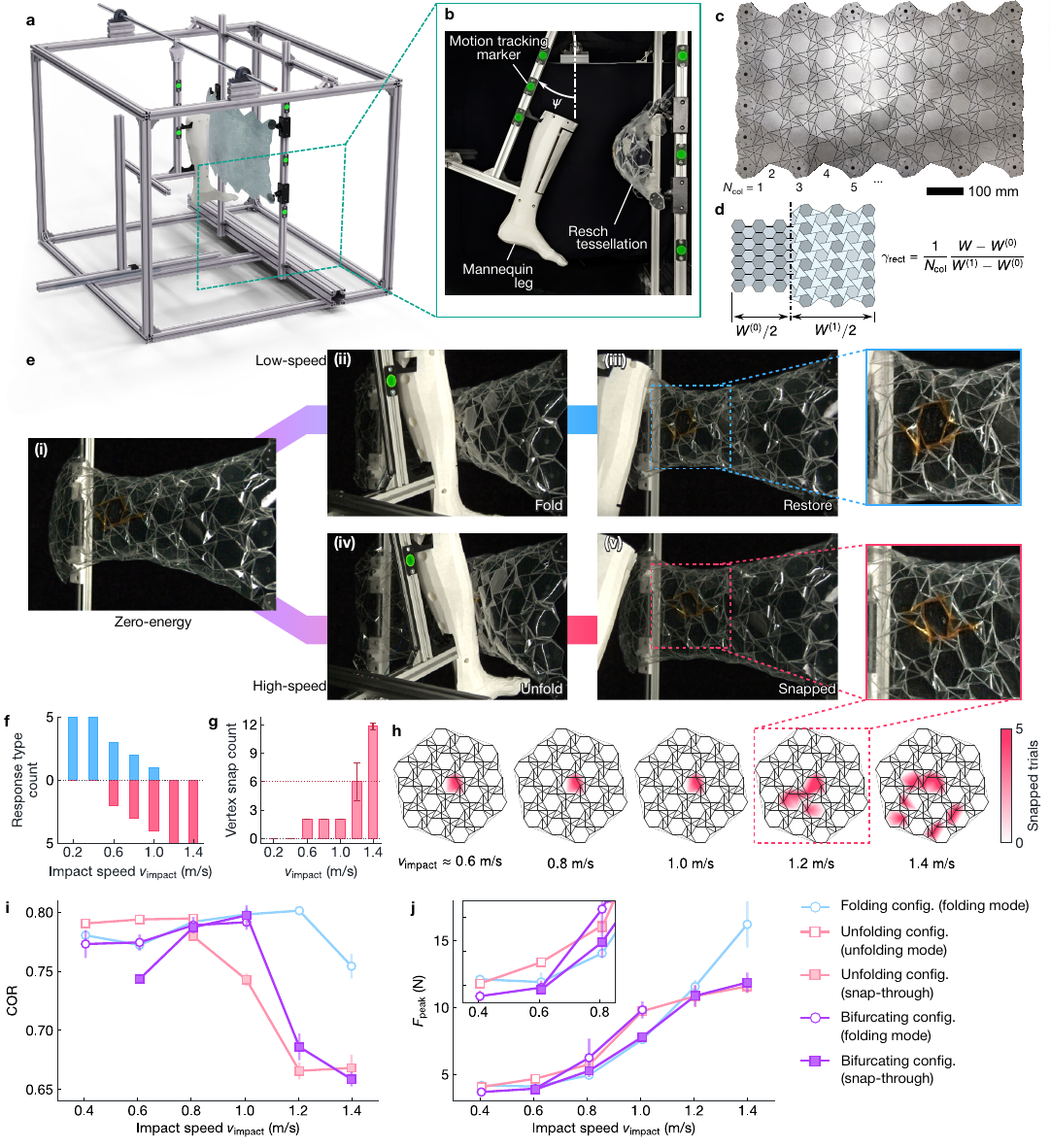}
    \caption[Large-scale Resch-patterned impact mitigation system.]{
    \textbf{Large-scale Resch-patterned impact mitigation system.}
    (\textbf{a}) Pendulum-based experimental set-up.
    (\textbf{b}) Magnified view of components and  pendulum angle $\psi$.
    (\textbf{c}) Laser-cut rectangular Resch origami tessellation (PET).
    (\textbf{d}) Definition of folding ratio $\gamma_{\mathrm{rect}}$.
    (\textbf{e}) Snapshot of Resch origami in bifurcating configuration at (top)  {$v_\mathrm{impact}=0.6$ m/s} ($\psi_0=15^\circ$) for low impact speed and (bottom) {$v_\mathrm{impact}=1.2$ m/s} ($\psi_0=30^\circ$) for high impact speed.
    {Magnified views of the central region are shown next to panels (iii) and (v) to highlight the restored and snapped configurations, respectively.}
    (\textbf{f}) Histogram of folding and snap-through as a function of impact speed. 
    (\textbf{g}) Average number of snapped vertices per trial (standard deviation shown as error bars).
    (\textbf{h}) Location of snapped vertices.
    (\textbf{i}) Coefficient of restitution.
    (\textbf{j}) Peak force measured by the force sensor embedded in the mannequin.
    }
    \label{fig:main_dynamic_pendulum}
\end{figure*}

We further assess the scalability and practical relevance of Resch origami by fabricating a larger bumper-like tessellation and emulating its pedestrian impact scenario (Fig.~{\ref{fig:main_dynamic_pendulum}}a).
The system comprises an adult-size mannequin leg mounted on a pendulum (Fig.~{\ref{fig:main_dynamic_pendulum}}b) and a rectangular Resch tessellation with $11\times6$ hexagons (Fig.~{\ref{fig:main_dynamic_pendulum}}c; Methods, Supplementary Note~8). 
To probe dynamic bifurcation and its onset with impact speed, we vary the initial pendulum angle $\psi_0$ in $5^\circ$ increments from $10^\circ$ to $35^\circ$~(Fig.~{\ref{fig:main_dynamic_pendulum}}b), {corresponding to impact speeds $v_\mathrm{impact}$ of approximately $0.2$ to $1.4$ m/s.} 
We prepare three configurations---folding, unfolding, and bifurcating---by adjusting the projected width of the tessellation (Fig.~{\ref{fig:main_dynamic_pendulum}}d), with narrower (wider) widths favoring folding (unfolding).
These correspond to folding ratios of $\gamma_\mathrm{rect} \approx 0.65$, $0.50$, and $0.44$, respectively.

By design, the folding and unfolding configurations exhibit their respective responses, including snap-through~(Supplementary Note~8).
The bifurcating configuration instead switches from folding to unfolding as the impact speed increases.
Figure~{\ref{fig:main_dynamic_pendulum}}e compares its deformation at {$v_\mathrm{impact}=0.6$ and $1.2$~m/s}.
At the lower speed, the center contracts and temporarily stores part of the mannequin's kinetic energy as recoverable strain energy.
The structure then returns to its convex shape and rebounds the mannequin.
At the higher speed, the center expands and follows the unfolding pathway.
This difference is evident in sub-panels~(ii) and~(iv): the distance between the upper and lower edges near the center decreases at low speed but increases at high speed.
After the mannequin swings back, the folding response fully recovers~(sub-panel~(iii)).
By contrast, the high-speed response remains in the second stable state~(sub-panel~(v)), {which can be mechanically reset by applying a reverse displacement~(Supplementary Notes~2.2 and~10.3)}.

Similar to the orbital tessellations, the folding-to-unfolding transition occurs gradually (Fig.~{\ref{fig:main_dynamic_pendulum}}f) over the range of critical {impact speeds between $v_\mathrm{impact}\approx0.6$ and $1.0$ m/s}.
Moreover, the number of snapped vertices increases and propagates farther from the center as the impact speed increases (Fig.~{\ref{fig:main_dynamic_pendulum}}g,~h), consistent with the observations in the multi-orbit tessellation cases (Supplementary Note~8).

{We assess energy-absorption behavior using the COR, which characterizes the associated loss of impactor kinetic energy~(Fig.~{\ref{fig:main_dynamic_pendulum}}i); a mass-normalized comparison using the specific energy absorption~(SEA) is additionally provided in Supplementary Note~6.2.
Impact-mitigation performance is evaluated using the peak impact force~(Fig.~{\ref{fig:main_dynamic_pendulum}}j), peak angular acceleration, and crush force efficiency~(CFE; Extended Data Fig.~5).}
For the folding configuration~(blue), the COR {remains nearly constant over a wide range of impact speeds}.
The unfolding configuration~(red) at low impact speeds~({$v_\mathrm{impact}\in[0.4,0.8]$ m/s}) remains near the first stable state, resulting in a higher COR than the folding configuration.
At higher impact speeds~({$v_\mathrm{impact}\in[0.8,1.4]$ m/s}), however, the unfolding configuration undergoes snap-through, leading to a pronounced reduction in COR relative to the folding configuration.
These trends indicate that folding absorbs more energy at low impact speeds, whereas unfolding, accompanied by snap-through, becomes more effective at higher impact speeds.

Remarkably, the bifurcating configuration combines the advantages of both modes.
At low impact speeds, it follows the folding response, yielding lower COR~(purple datasets collapsing onto blue).
At higher speeds, it switches to the unfolding mode and similarly achieves reduced COR~(purple approaching red).
A consistent trend is observed in the peak force (Fig.~{\ref{fig:main_dynamic_pendulum}}j), {peak acceleration, and CFE (Extended Data Fig.~{\ref{fig:ex_rectangular_PET_performance}})}, where the bifurcating configuration transitions between the lower-force regimes of the folding and unfolding modes.
Collectively, these results demonstrate that the Resch origami adaptively selects the deformation mode that maximizes impact mitigation performance.

\section*{Conclusion}

In summary, our findings reveal that dynamic bifurcation in a bistable Resch origami structure enables a fully passive, self-adaptive response for impact mitigation.
Unlike conventional protective strategies tuned to a specific impact condition, the intrinsic geometric duality of the Resch pattern allows the structure to select its deformation pathway according to the impact speed relative to a tunable critical threshold.
Below this threshold, the structure follows a compliant, monostable folding pathway; above it, the structure transitions to the bistable unfolding pathway and undergoes snap-through, enhancing energy dissipation.
{We validate this geometric mechanism across different tessellation sizes and constituent materials, including PET and stainless steel, and further show that the self-adaptive mode switching persists under oblique impacts~(Extended Data Fig.~{\ref{fig:si_setup_pendulum_meter_oblique}})}.
{The numerical parametric analysis further identifies impactor mass, contact size, structural stiffness, and initial folding ratio as key factors governing the critical speed and spatial extent of snap-through.}
{By linking the quasi-static bifurcation landscape to impact-speed-dependent pathway selection, this work provides a geometric framework for designing passive structures that adapt their impact response without external sensing or actuation.}

%TC:ignore
\clearpage
\section*{Methods}

\subsection*{Dynamic bar--hinge--mass numerical model}
We employ an origami model consisting of an axial bar, torsional hinge, and lumped mass in this study.
The resulting equation of motion of the $i$-th vertex~(i.e., mass) of the origami is
$m_i\ddot{\mathbf{u}}_i+\mathbf{F}_{i,\rm bar}+\mathbf{F}_{i,\rm tor}+\mathbf{F}_{i,{\rm damp}}=\mathbf{F}_{\rm ext},$
where $m_i$ is the lumped mass, $\mathbf{u}_i$ is the displacement vector, $\mathbf{F}_{i,\rm bar}$ and $\mathbf{F}_{i,\rm tor}$ are the internal forces due to axial spring and torsional spring deformation, respectively.
The axial bar stiffness is estimated based on the material properties~\mbox{\cite{Liu2017}}, and the torsional spring stiffnesses are calibrated through a single hinge compression test.
Energy dissipation is governed by the $\mathbf{F}_{i,\rm damp}$ term.
The sum of these terms balances the external force $\mathbf{F}_{\rm ext}$.
This term can be zero for free unfolding simulation, gradually increasing over time for examining quasi-static response, or a contact force with another object for the dynamic impact cases.
The complete formulation, including the stiffness calibration and damping model, is
provided in Supplementary Note 4.

Building on established bar--hinge--mass models~\cite{Liu2017, Zhu2019, Yu2023}, we incorporate gravitational and viscoelastic contact forces to simulate impact-induced deformation on a millisecond timescale.
The model is implemented using an in-house code written in Python and Fortran.
To resolve the short-duration dynamics while controlling the integration error, we use an adaptive eighth-order Dormand--Prince method with embedded fifth- and third-order error estimators~{\mbox{\cite{Hairer1993}}}.
The maximum time-step size is $\Delta t=10^{-5}$ s, and both the relative and absolute error tolerances are set to $\epsilon=10^{-12}$.
All computations use double-precision floating-point arithmetic.
The numerical parameters are provided in Supplementary Note~14 and Supplementary Tables~2,3, and the computational algorithm is described in Supplementary Note~15.

\subsection*{Quasi-static folding motion and bifurcation diagram}
{
The folding motion shown in Fig.~{\ref{fig:main_schematic}}a--c is obtained by a damped dynamical-relaxation simulation.
Starting from the fully folded configuration, the centroid of the central hexagon is fixed, while all other degrees of freedom remain unconstrained.
Viscous damping is applied to suppress transient dynamics as the structure deploys toward the fully deployed configuration, thereby tracing the quasi-static, quasi-rigid folding motion.
}
{
To construct the strain-energy landscape and bifurcation diagram in Fig.~{\ref{fig:main_schematic}}d, we use two prescribed deformation coordinates: the normalized out-of-plane displacement $u/a$ of the central hexagon and the folding ratio $\gamma$.
These coordinates are independently varied over a regular $(u/a,\gamma)$ grid.
For each prescribed pair, all remaining degrees of freedom are relaxed using the same damped dynamical-relaxation procedure.
The plotted energy is the change in strain energy relative to the bifurcating configuration (iii),}
\begin{align}
\Delta U_{\rm total} =
\left(U_{\rm bar}+U_{\rm tor}\right)
-
\left(U_{\rm bar}^{\rm (iii)}+U_{\rm tor}^{\rm (iii)}\right),
\end{align}
{where $U_{\rm bar}$ and $U_{\rm tor}$ are the axial-bar and torsional-spring strain energies, respectively.
Thus, the landscape represents the mechanically stored energy over the constrained deformation space.
The two valley paths emerging from configuration (iii) correspond to the folding and unfolding deformation modes observed in Fig.~{\ref{fig:main_schematic}}e,f.
}

\subsection*{Fabrication}

Orbital Resch origami prototypes are laser-cut from $0.254$-mm-thick PET sheets and manually folded according to the crease pattern shown in Supplementary Fig.~1.
The initial configuration of each prototype is set by heat treatment in a convection oven at $80\,^\circ\mathrm{C}$ for three hours using custom 3D-printed supports designed from the simulated Resch geometry.
For the pendulum tests, an approximately $750\times450$ mm PET sheet is laser-cut and manually folded, resulting in projected dimensions of approximately $550\times350$ mm.
The EPS specimens are shaped manually using a hot wire.
The EPS block is first cut to a thickness of approximately $18$ mm and then trimmed to match the fully deployed profile of the orbital Resch specimen used in the drop tests.
Further details of the PET and EPS specimens are provided in Supplementary Notes~1, 7, and~8.

{For the stainless-steel specimens, the greater resistance of the sheet to folding requires a systematic examination of crease perforation and bridge designs.
We select designs that reproduce the intended folding and unfolding motions while maintaining durability under repeated deformation.
The selected crease patterns are laser-cut into $0.2$-mm-thick 304H stainless-steel sheets and manually folded to fabricate the specimens used in the impact experiments~(Supplementary Video~5).
For scalable production, the forming process must accommodate the biaxial contraction and facet rotation intrinsic to Resch folding.
We therefore develop a progressive, multi-stage stamping process that forms a flat sheet using a sequence of molds with increasing folding ratios.
We demonstrate this process using a larger paper tessellation and a $0.1$-mm-thick SUS304 specimen~(Extended Data Fig.~{\ref{fig:ex_mold_fabrication}}).
Further details of the metallic crease design, prototype fabrication, and stamping process are provided in Supplementary Notes~9--11.}

\subsection*{Experiments}

We characterize the orbital Resch origami using quasi-static compression and dynamic drop tests.
In both experiments, the central hexagon is constrained against in-plane translation at its centroid but is allowed to move in the out-of-plane direction and rotate during deformation.
{Three equally spaced outer hexagons are connected through ball joints to radial sliders, allowing them to translate radially and rotate about their centroids.}
For the quasi-static tests, a linear stage applies a controlled out-of-plane displacement to the central hexagon, while a load cell and laser sensor measure the resulting force and displacement, respectively.

{For the single-orbit drop tests, a hemispherical impactor with a diameter of $D_\mathrm{impactor}=32$ mm is released along a vertical shaft onto the central hexagon using the same radially guided mounting system.
Two high-speed cameras record the impact event: a front-facing camera captures the impactor trajectory, while a close-up camera resolves the local deformation of the Resch origami, as shown in Fig.~{\ref{fig:main_dynamic_bifurcation}}f,g.
Markers attached to the impactor are tracked using digital image correlation~(DIC) to obtain its displacement and velocity.}

The large-scale rectangular tessellation is evaluated using a pendulum system designed to emulate pedestrian impact.
{The tessellation is connected to the supporting frame through ball joints at its four corner hexagons, fixing the attachment points in translation while permitting rotation.
An adult-sized mannequin leg mounted on the impact pendulum strikes the center of the tessellation, and force transducers attached to the leg measure the contact force.
Two high-speed cameras record the global pendulum motion and the local tessellation deformation.
Markers on the pendulum are tracked using DIC to reconstruct its angular motion and determine the impact speed and angular acceleration.}
Detailed descriptions of the experimental components, assembly procedures, and boundary conditions are provided in Supplementary Notes~2, 3, 7, and~8.

\section*{Acknowledgments}
We acknowledge the support from Toyota Research Institute North America, the Air Force Office of Scientific Research (FA2386-24-1-4051), SNU-IAMD, SNU-IOER, and the National Research Foundation of Korea [2023R1A2C2003705 and 2022H1D3A2A03096579].
We also thank Mr. Yasuhito Miyazawa for his insightful comments on the experimental setup design.

\section*{Author contributions}
Y.M. and J.Y. conceived the project;
Y.M., Y.S., S.S., U.G., and J.Y. conceptualized the project;
Y.M. and C.-Y.C. developed the theoretical framework;
Y.M., R.T.A., M.C., D.L., and S.K. fabricated the specimens;
Y.M., Q.L., R.T.A., and D.L. designed the experimental setup;
Y.M., Q.L., R.T.A., K.Y., M.C., and D.L. performed experiment;
Y.M. analyzed experimental results, developed the software code, and carried out numerical simulation;
Y.M. and J.Y. wrote the paper;
J.Y. supervised the project.

\section*{Competing interests}
The authors declare no competing interests.

\clearpage

\clearpage
\setcounter{section}{0}
\setcounter{equation}{0}
\setcounter{figure}{0}
\setcounter{table}{0}
\renewcommand{\figurename}{Extended Data Fig.}

\begin{figure*}[htbp]
    \centering
    \includegraphics[width=183truemm]{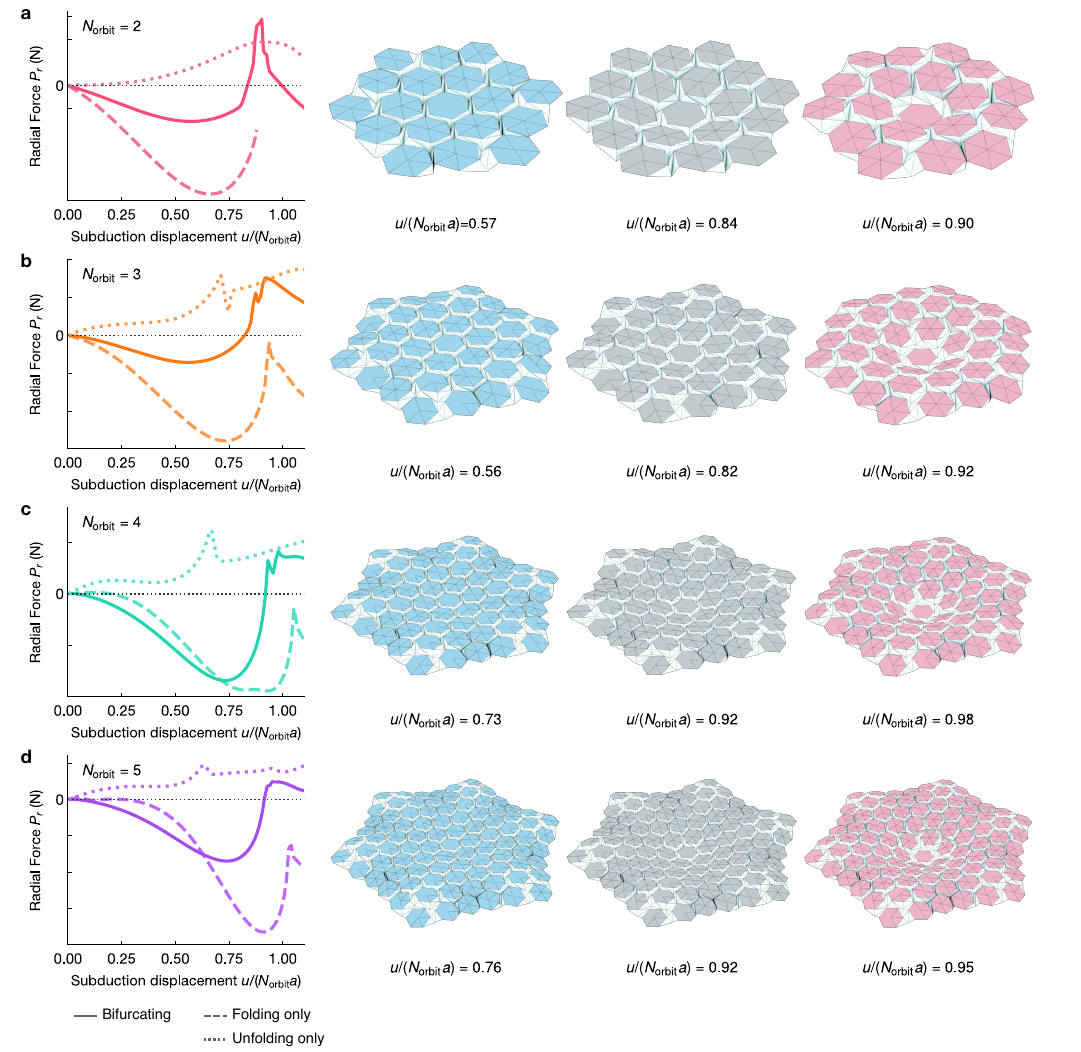}
    \caption[Radial force of multi-orbit tessellations.]
    {{\textbf{Radial force of multi-orbit tessellations.}
    Radial forces evaluated at the outermost orbit during quasi-static subduction for
    (\textbf{a}) $N_\mathrm{orbit}=2$ at $\gamma=\{0.4,0.6,0.8\}$,
    (\textbf{b}) $N_\mathrm{orbit}=3$ at $\gamma=\{0.4,0.5,0.6\}$,
    (\textbf{c}) $N_\mathrm{orbit}=4$ at $\gamma=\{0.4,0.5,0.6\}$,
    (\textbf{d}) $N_\mathrm{orbit}=5$ at $\gamma=\{0.4,0.5,0.6\}$.
    Around the bifurcating configuration $\gamma$ (solid lines), the radial force initially indicates a favor for folding motion (negative force), and switches to unfolding (positive force) and eventually snap-through at the discontinuity near the maximum force.
    On the other hand, for $\gamma$ less (larger) than the bifurcating $\gamma$, the radial force remains positive (negative) throughout subduction, indicating that the Resch origami remains in unfolding (folding) mode.
    The subduction displacement is normalized by the orbit number $N_\mathrm{orbit}$ for comparison.}
    }
    \label{fig:ex_radial_force_multiorbit}
\end{figure*}

\begin{figure*}[htbp]
    \centering
    \includegraphics[width=\linewidth]{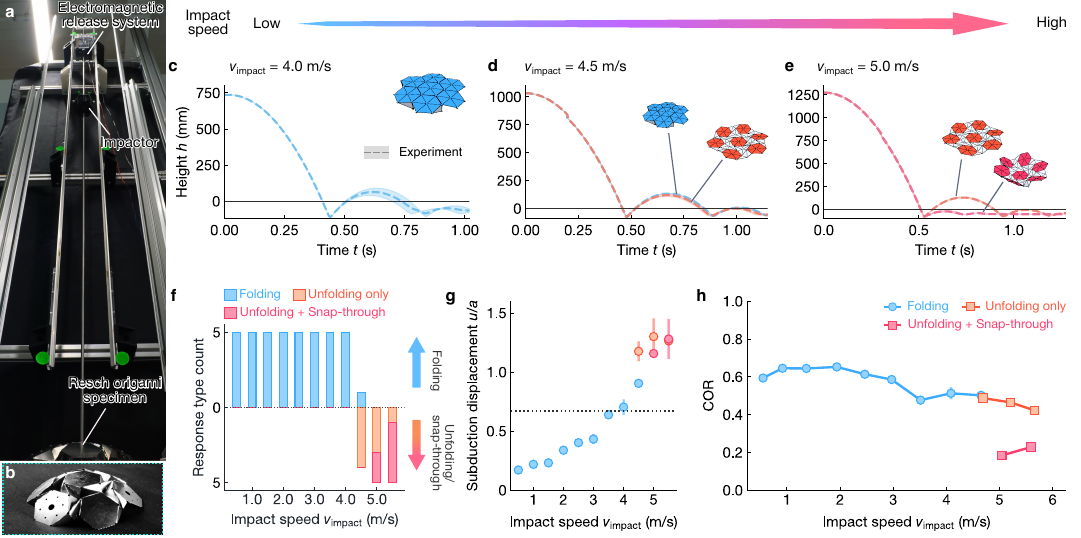}
    \caption[Dynamic bifurcation of stainless-steel Resch origami.]
    {{\textbf{Dynamic bifurcation of stainless-steel Resch origami.}
    (\textbf{a}) Drop tower configuration.
    A $135$~g impactor installed on the higher tower is used to access the higher impact energy regime of the stainless-steel specimen.
    (\textbf{b}) Stainless-steel Resch prototype.
    The height of the impactor at
    (\textbf{c}) $v_{\rm impact}=4.0$ m/s,
    (\textbf{d}) $4.5$ m/s, and
    (\textbf{e}) $5.0$ m/s.
    Dashed lines, experimental mean; shaded regions, standard deviation over five trials.
    The inset figures of each panel schematically display the folding mode that Resch origami follows.
    (\textbf{f}) Histogram of folding, unfolding, and snap-through as a function of impact speed.
    (\textbf{g}) Subduction displacement as a function of impact speed.
    (\textbf{h}) Coefficient of restitution as a function of impact speed.
    Blue circle symbol, folding mode;
    orange square symbol, unfolding mode;
    red square symbol, snap-through mode.}
    }
    \label{fig:ex_dynamic_bifurcation_stainless}
\end{figure*}

\begin{figure*}[htb]
    \centering
    \includegraphics[width=183truemm]{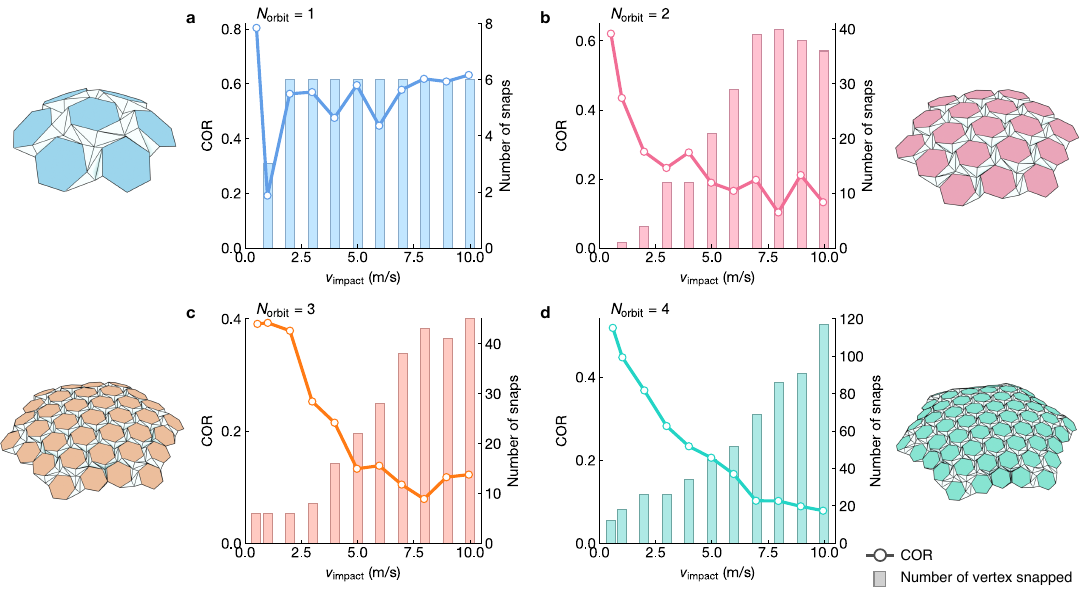}
    \caption[Energy absorption performance and vertex snap count.]{{\textbf{Energy absorption performance and vertex snap count.}
    Coefficient of restitution $\mathrm{COR}$ and number of snapped vertices across the tested impact speed $v_\mathrm{impact}$ for
    (\textbf{a}) $N_\mathrm{orbit}=1$,
    (\textbf{b}) $N_\mathrm{orbit}=2$,
    (\textbf{c}) $N_\mathrm{orbit}=3$, and
    (\textbf{d}) $N_\mathrm{orbit}=4$.}
    }
    \label{fig:ex_parametric_vertex_snap_count_cor}
\end{figure*}

\begin{figure*}[htbp]
    \centering
    \includegraphics[width=\linewidth]{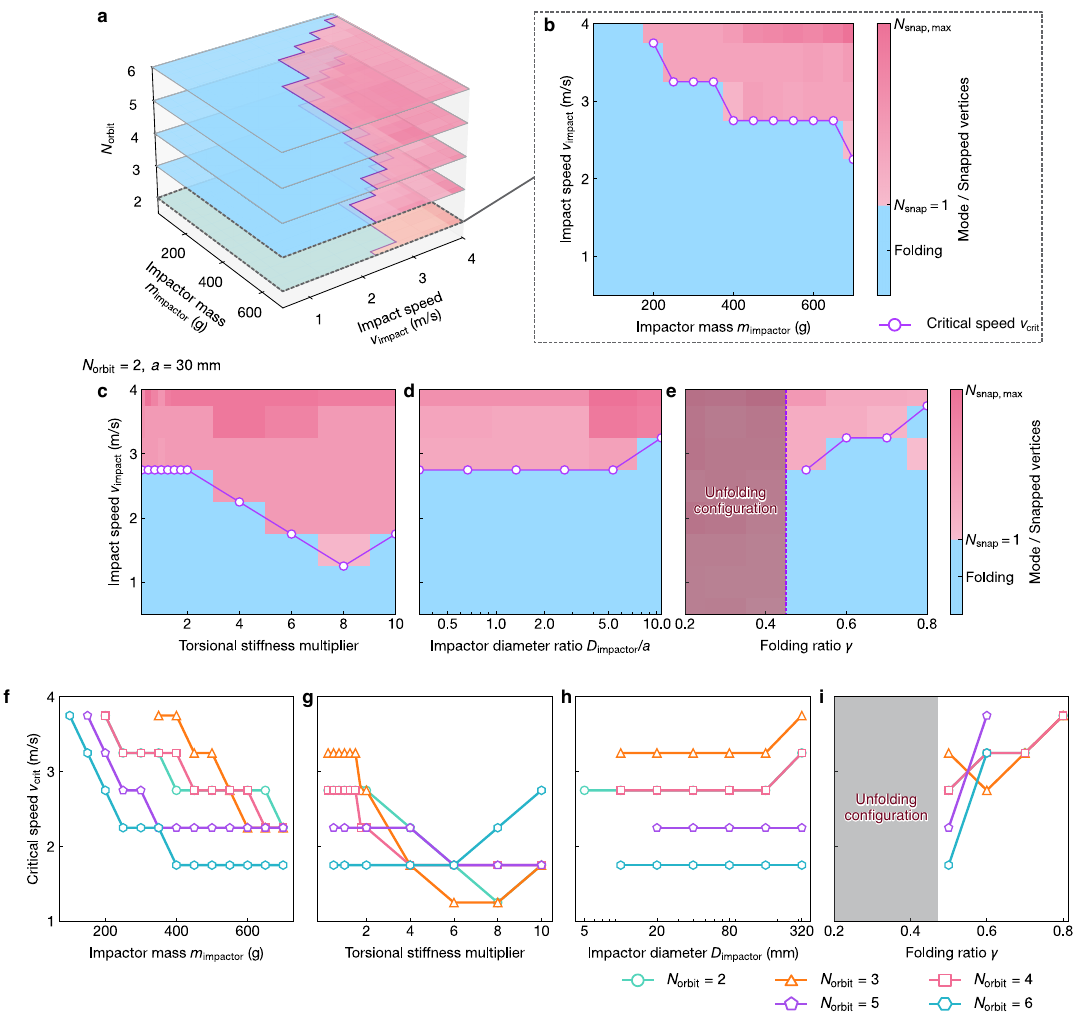}
    
    \caption[Parameter space governing dynamic bifurcation.]{{
    \textbf{Parameter space governing dynamic bifurcation.}
    (\textbf{a}) Stacked deformation-mode maps in the impact-speed--impactor-mass space for tessellations with $N_\mathrm{orbit}=2$--6.
    (\textbf{b}--\textbf{e}) Deformation-mode maps for $N_\mathrm{orbit}=2$ as functions of impact speed $v_\mathrm{impact}$ and
    (\textbf{b}) impactor mass $m_\mathrm{impactor}$,
    (\textbf{c}) torsional-stiffness multiplier $\kappa$,
    (\textbf{d}) impactor diameter ratio $D_\mathrm{impactor}/a$, and
    (\textbf{e}) initial folding ratio $\gamma$.
    Blue regions indicate folding without snap-through, whereas the pink shading indicates the number of snapped vertices $N_\mathrm{snap}$.
    Purple markers and lines denote the critical speed $v_\mathrm{crit}$ for the onset of snap-through.
    (\textbf{f}--\textbf{i}) Critical speed $v_\mathrm{crit}$ for $N_\mathrm{orbit}=2$--6 as a function of
    (\textbf{f}) impactor mass,
    (\textbf{g}) torsional-stiffness multiplier,
    (\textbf{h}) impactor diameter, and
    (\textbf{i}) initial folding ratio.
    Unless specified, the baseline parameters are
    $m_\mathrm{impactor}=500$ g,
    $\kappa=0.5$,
    $D_\mathrm{impactor}=40$ mm, and
    $\gamma=0.5$; the facet edge length is $a=30$ mm.
    The shaded regions in (\textbf{e, i}) correspond to initial geometries in the unfolding configuration.
    }}
    \label{fig:ex_parametric_sweep}
\end{figure*}

\begin{figure*}[htbp]
    \centering
    \includegraphics[width=\linewidth]{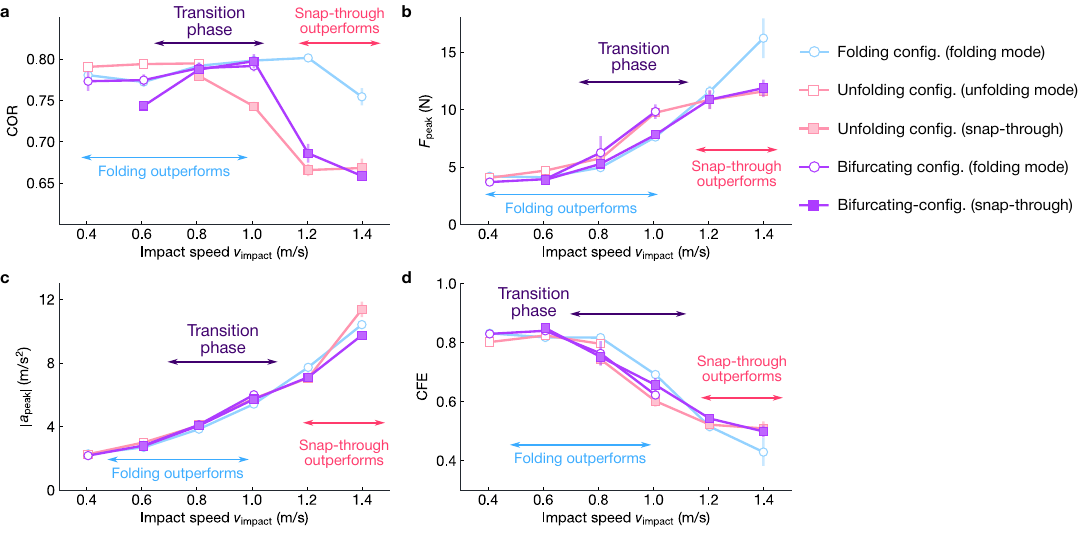}
    \caption[Performance metrics and mode transition in pendulum test.]{{
    \textbf{Performance metrics and mode transition in pendulum test.}
    (\textbf{a}) Coefficient of restitution of the mannequin pendulum.
    (\textbf{b}) Peak force experienced by the mannequin leg upon impact.
    (\textbf{c}) Peak acceleration experienced by the mannequin leg.
    (\textbf{d}) Crushing force efficiency based on force sensor measurement.
    The CFE is defined as $\dfrac{\overline{F}_{\mathrm{impact}}}{F_{\mathrm{peak}}}$, where $\overline{F}_{\mathrm{impact}}=\dfrac{1}{u_{\mathrm{indent}}}\displaystyle\int_{0}^{u_{\mathrm{indent}}}F(u)\,du$ is the mean impact force over the indentation displacement $u_{\mathrm{indent}}$.}
    }
    \label{fig:ex_rectangular_PET_performance}
\end{figure*}

\begin{figure*}[h]
    \centering
    \includegraphics[width=183truemm]{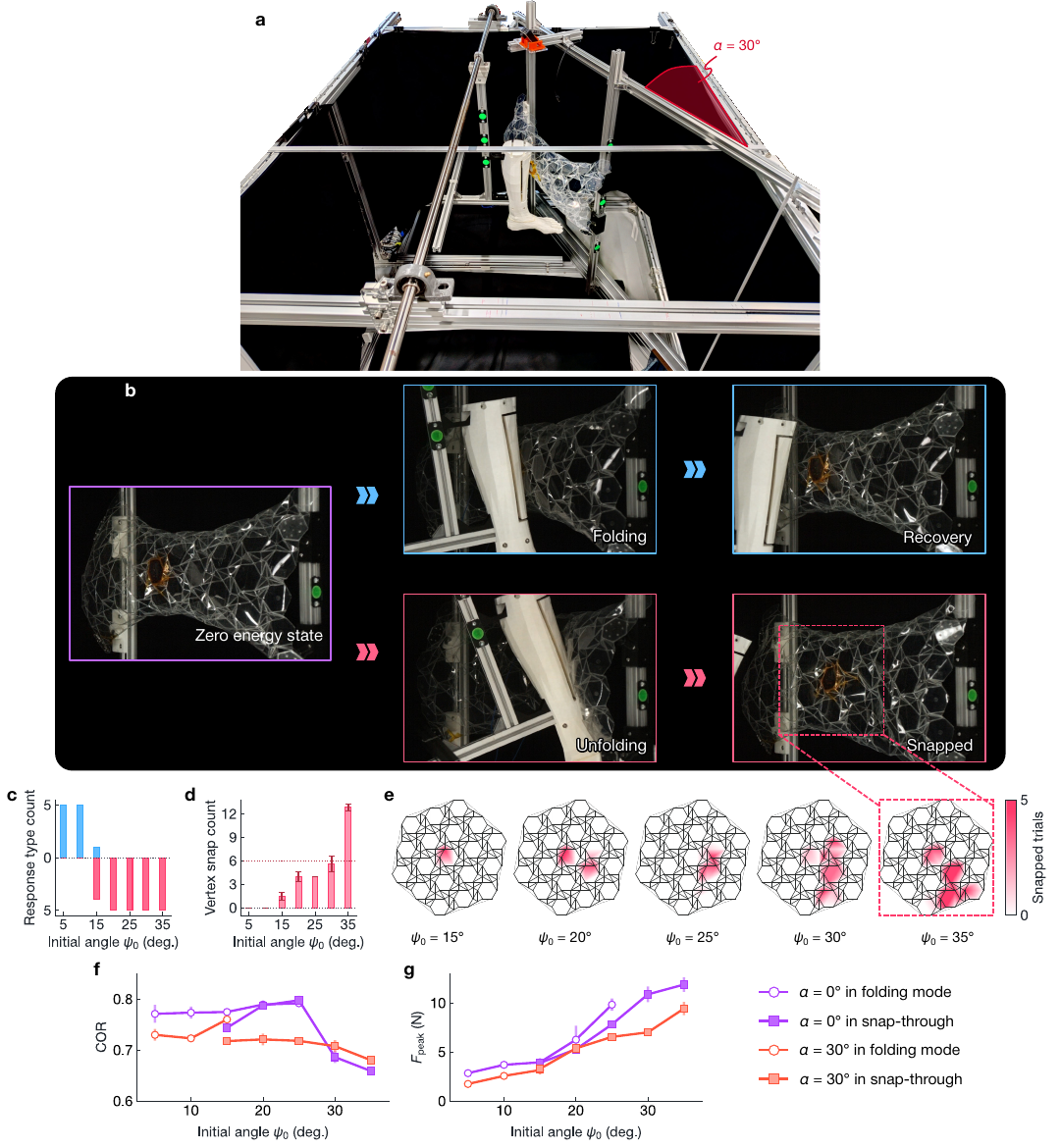}
    \caption[Oblique pendulum impact test.]{\textbf{Pendulum setup for oblique impact test.}
    (\textbf{a}) Overview of the experimental setup.
    (\textbf{b}) Snapshots of Resch origami in a bifurcating configuration under impact at low~(top; $\psi_0=10^\circ$) and high~(bottom; $\psi_0=35^\circ$) speed.
    (\textbf{c}) Response type count.
    (\textbf{d}) Snapped vertex count.
    (\textbf{e}) Snapped vertex distribution.
    (\textbf{f}) Coefficient of restitution.
    (\textbf{g}) Peak impact force.
    }
    \label{fig:si_setup_pendulum_meter_oblique}
\end{figure*}

\begin{figure*}[htbp]
    \centering
    \includegraphics[width=\linewidth]{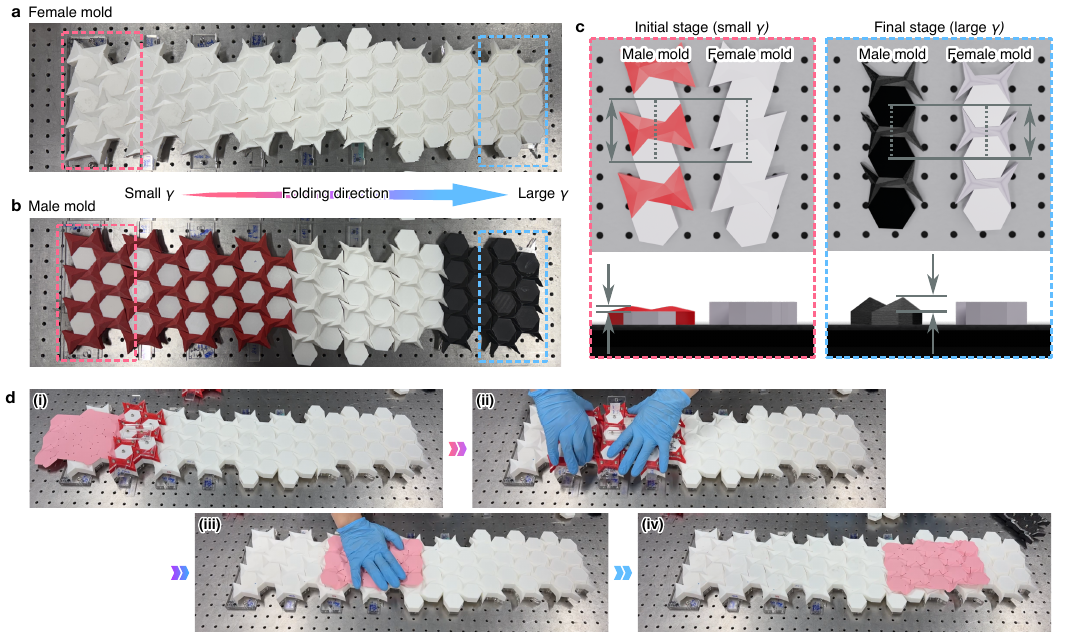}
    \caption[Mold stamping fabrication of Resch origami tessellation.]{{
    \textbf{Mold stamping fabrication of Resch origami tessellation.}
    (\textbf{a},\textbf{b}) Gradual step female and male molds.
    (\textbf{c}) Initial and final steps with different mold geometry.
    With a large folding ratio $\gamma$, the hexagons are closer to each other~(top), with the startuck being deeper (i.e., higher male mold height; bottom).
    (\textbf{d}) Step-by-step mold stamping using a sheet of paper.
    A Resch origami is gradually folded as the sheet progresses from right to left (small to large $\gamma$).
    }
    }
    \label{fig:ex_mold_fabrication}
\end{figure*}

%TC:endignore

\end{document}